\documentclass[12pt]{iopart}
\usepackage{epsf}
\begin{document}

\def\dj{\leavevmode\setbox0=\hbox{d}\dimen0=\wd0
        \setbox0=\hbox{d}\advance\dimen0 by -\wd0
        \rlap{d}\kern\dimen0\hbox to \wd0{\hss\accent'26}}

\title{Three-centre cluster structure in $^{11}$C and $^{11}$B}

\author{N Soi\'{c}\dag \ddag , M Freer\ddag , L Donadille\ddag,
N M Clarke\ddag, P J Leask\ddag, W N Catford\S, K L Jones\S,
D Mahboub\S, B R Fulton$\|$, B J Greenhalgh$\|$,
D L Watson$\|$ and D C Weisser\P }

\address{\dag Ru\dj er Bo\v{s}kovi\'{c} Institute, Bijeni\v{c}ka 54,
HR-10000 Zagreb, Croatia}
\address{\ddag School of Physics and Astronomy, University of Birmingham,
 Edgbaston, Birmingham B15 2TT, United Kingdom}
\address{\S School of Electronics and Physical Sciences, University of
Surrey, Guildford GU2 7XH, United Kingdom} 
\address{ $\|$ Department of Physics, University of York, Heslington, York,
 YO10 5DD, United Kingdom} 
\address{ \P Department of Nuclear Physics, The Australian National University,
Canberra ACT 0200, Australia}

\begin{abstract}

Studies of the $^{16}$O($^{9}$Be,$\alpha$$^{7}$Be)$^{14}$C,
$^{7}$Li($^{9}$Be,$\alpha$$^{7}$Li)$^{5}$He and
$^{7}$Li($^{9}$Be,$\alpha$$\alpha$t)$^{5}$He reactions at
E$_{beam}$=70 and 55 MeV have been performed using resonant particle
spectroscopy techniques. The $^{11}$C excited states decaying into
$\alpha$+$^{7}$Be(gs) are observed between 8.5 and 13.5 MeV.
The $\alpha$+$^{7}$Li(gs), $\alpha$+$^{7}$Li*(4.652 MeV) and
t+$^{8}$Be(gs) decays of $^{11}$B excited states between 9 and 19 MeV
are observed. The decay processes are used to indicate the possible 
three-centre 2$\alpha$+$^{3}$He (2$\alpha$+$^{3}$H) cluster structure of 
observed states. This cluster structure is more prominent in 
the positive-parity states, where two rotational bands with large 
deformations are suggested. Excitations of some of the observed 
T=1/2 resonances coincide with the energies of previously
measured T=3/2 isobaric analogs of the $^{11}$Be states, 
indicating that these states may have mixed isospin.

\end{abstract}

%\pacs{24.30.Gd, 25.70.Ef, 27.20.+n}


%\maketitle

%\section{Introduction}

 Many light nuclei possess well developed cluster structure made of
$\alpha$-particles as main building block. The best known examples are
2$\alpha$ and 3$\alpha$ cluster structures of $^{8}$Be and $^{12}$C.
In recent years attention has been focused on neutron-rich Be and C nuclei,
where evidence for developed cluster structures was found (for example
\cite{mi,freerpr,mil02,so13c14c} and references therein). 
It is interesting to investigate influence of $\alpha$-clustering on
structure of the neutron deficient $^{11}$C nucleus and its mirror 
nucleus $^{11}$B for which the experimental evidence for cluster 
structures is scarce. Here we present results of the experimental 
studies which probe cluster structure of $^{11}$C and $^{11}$B via 
the $\alpha$-decay of their excited states and for $^{11}$B also via 
2$\alpha$+t decay.

%\section{Results}

\begin{figure}
\begin{center}
\epsfxsize=0.8\textwidth
\epsfbox{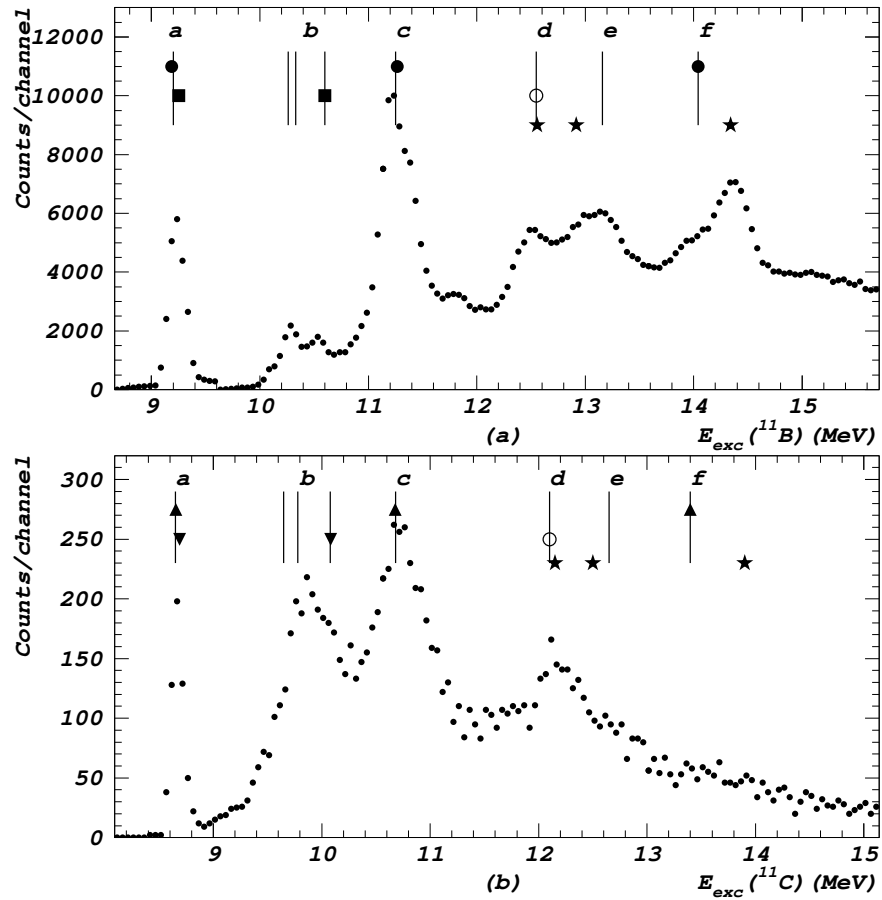}
\end{center}
\caption{\label{11c11b} $^{11}$B excitation energy spectrum from the
$^{7}$Li($^{9}$Be,$\alpha$$^{7}$Li)$^{5}$He reaction (a) and $^{11}$C
excitation energy spectrum from the $^{16}$O($^{9}$Be,$\alpha$$^{7}$Be)$^{14}$C
reaction (b) at E$_{beam}$=70 MeV. Spectra are shifted so that the 
7/2$^{+}$,5/2$^{+}$ doublets are aligned. Lines mark positions of the 
states with the same value of the spin and parity populated in both nuclei. 
The positions of the K=5/2$^{+}$ band members are marked with full circles 
and triangles, K=3/2$^{+}$ band members with full squares and inverted 
triangles, possible 9/2$^{+}$ states in K=3/2$^{+}$ bands with open circles. 
The excitations of the isobaric analogues of the lowest three $^{11}$Be states 
are marked with stars.   }
\end{figure}

The measurements were performed at the Australian National University's
14UD tandem accelerator facility. 70 MeV and 55 MeV $^9$Be beam were incident
on a thin Li$_2$O$_3$ foil. Reaction products were detected in an array of
four charged-particle telescopes which provided charge and mass resolution
up to Be. Telescopes 1 and 2 were located with their centres at 
$\sim$17.5$^{\circ}$ from the beam axis and telescopes 3 and 4 at 
$\sim$29$^{\circ}$. The measurement of the energies and angles of detected 
particles permitted the kinematics of the many-body reactions to be fully 
reconstructed. Further experimental details can be found in 
\cite{so13c14c,so11c11b}.

\begin{figure}
\begin{center}
\epsfxsize=1.\textwidth
\epsfbox{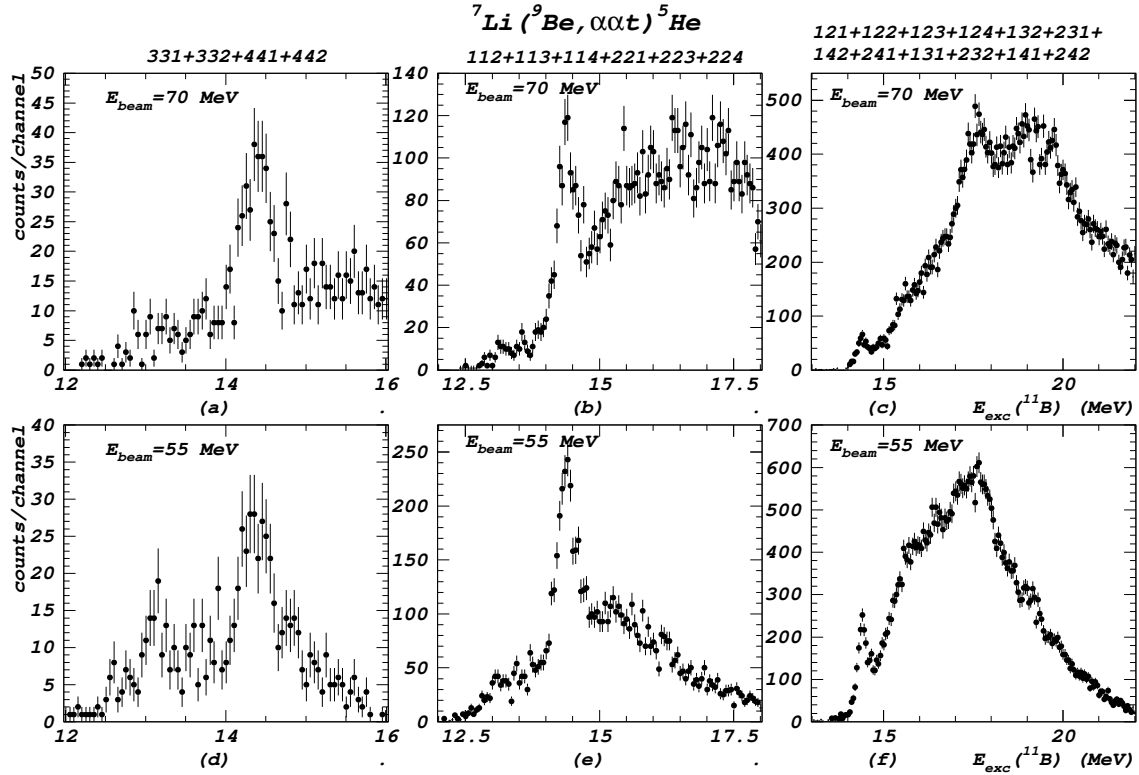}
\end{center}
\caption{\label{aat} $^{11}$B excitation energy spectra from the
$^{7}$Li($^{9}$Be,$\alpha$$\alpha$t)$^{5}$He reaction at E$_{beam}$=70 and
55 MeV. Numbers at the top denote detectors combination in which events were
detected: (ijk) i,j are for $\alpha$-particles and k for triton.  }
\end{figure}

The $\alpha$-decay of excited states in $^{11}$B has been studied using the
$^{7}$Li($^{9}$Be,$^{11}$B* $\rightarrow$ $\alpha$+$^{7}$Li)$^{5}$He
reaction (Q=-2.4 MeV). The $^{11}$B excitation energy spectra show resonances 
at 9.2, 10.3, 10.55, 11.2, (11.4), 11.8, 12.5, (13.0), 13.1, (14.0), 14.35, 
(17.4) and (18.6) MeV. A study of the $\alpha$-decay of $^{11}$C excited 
states has been performed using the 
$^{16}$O($^{9}$Be,$^{11}$C* $\rightarrow$ $\alpha$+$^{7}$Be)$^{14}$C reaction 
(Q=-14.602 MeV). The $^{11}$C excitation energy spectra provide evidence for 
peaks at 8.65, 9.85, 10.7 and 12.1 MeV and indications for peaks at 12.6 
and 13.4 MeV. Detailed description and discussion of the results are presented 
in \cite{so11c11b}. Analysis of these data shows that main reaction mechanism 
for the population of the $\alpha$-decaying states in both nuclei was 
two-nucleon 
transfer onto the cluster nucleus $^{9}$Be. Figure \ref{11c11b} presents 
comparison of the $^{11}$B and $^{11}$C excitation energy spectra from these 
measurements. We observe the same series of excited states at the lower 
excitations in both nuclei: unresolved doublet of 7/2$^+$, 5/2$^+$ states 
(marked 'a'), a 3/2$^-$, 5/2$^-$, 7/2$^+$ triplet ('b') which is $\sim$1.2 MeV 
above the doublet, 9/2$^+$ state ('c') at an excitation of 2 MeV higher than 
the doublet, then the state which is believed to be the isobaric analogue state
of the $^{11}$Be ground state ('d') which is $\sim$3.4 MeV above the doublet, 
and then weak states, which are probably 7/2$^+$ ('e') and 11/2$^+$ ('f'), 
and which are 3.9 and 4.8 MeV above the doublet's excitation. An interesting 
feature of these results is the observation of 
the $\alpha$-decay, which means isospin T=1/2, of the excited states proposed 
to be the isobaric analogue states of the $^{11}$Be states, which have T=3/2, 
in both nuclei. If the observed states are indeed those identified with T=3/2 
character, then our results show that the lowest three T=3/2 levels in $^{11}$B
and the first T=3/2 level in $^{11}$C probably have large isospin mixing.
Alternatively, the peaks may have a genuine T=1/2 character and may be linked 
to rotational bands. The present results show that six observed states in
$^{11}$B and six in $^{11}$C can be grouped in two rotational bands 
\cite{so11c11b}. Rotational bands with K=5/2$^+$ have been proposed beginning 
at 7.286 and 6.905 MeV in $^{11}$B and $^{11}$C respectively, with rotational 
members at 9.185 and 8.655 (7/2$^+$), 11.265 and 10.679 (9/2$^+$) and 14.04 
and 13.4 MeV (11/2$^+$). The moment of inertia I of these bands is very large,
with a rotational parameter $\hbar^{2}$/2I of 0.25 MeV which would correspond 
to very deformed structure. Other possible positive-parity bands in both 
$^{11}$B and 
$^{11}$C have been proposed with K=3/2$^+$, beginning at excitation energies
7.97784 and 7.4997 MeV, respectively. Rotational members are at 9.274 and 8.699
MeV (5/2$^+$) and 10.597 and 10.083 MeV (7/2$^+$) and the 9/2$^+$ members of 
these bands would be at 12.6 MeV in $^{11}$B and 12.1 MeV in $^{11}$C,
very close to the excitations of the proposed 1/2$^+$ T=3/2 states and
exactly at excitations where resonances are in the present spectra.
The rotational parameter $\hbar^{2}$/2I of these two bands would be 0.215 MeV.

Triple coincidence data provide clear evidence for the 
$^{7}$Li($^{9}$Be,$\alpha$$\alpha$t)$^{5}$He reaction (Q=-4.9 MeV) 
at both E$_{beam}$=70 
and 55 MeV \cite{soaat}. Reconstructed relative energy spectra for all 
possible combinations of 2 and 3 particles in the exit channel provide 
evidence for decays of $^{11}$B excited states into 2$\alpha$+t, the $^{8}$Be 
ground state and the first excited state into 2$\alpha$ and for $\alpha$+t 
decay of $^{7}$Li states at 4.652 and 6.6 MeV. The $^{11}$B excitation energy 
spectra reconstructed from the energies and momenta of all three detected 
decay 
products presented in Figure \ref{aat} show peaks at 13.1, 14.4 and 17.5 MeV. 
All three states are also observed in the $\alpha$+$^{7}$Li data. An analysis 
gives evidence for the t+$^{8}$Be(gs) decay of the $^{11}$B states at 13.1 and 
14.4 MeV and decays via $^{8}$Be(3.03 MeV) and $^{7}$Li(4.652 MeV) states of 
the 14.4 and 17.5 MeV states \cite{soaat}. These results are the first direct 
evidence for decay of $^{11}$B excited states into these three-body channels.

 Given the nature of the reaction processes, two-nucleon transfer onto the 
2$\alpha$+n nucleus $^{9}$Be, the observed strong $\alpha$-decay of these 
mirror states at excitations where various decay channels are open and known 
$\alpha$+t($^{3}$He) structure of $^{7}$Li($^{7}$Be), as well as observed 
decays of $^{11}$B states into 2$\alpha$+t, it is possible that these states
possess the three-centre 2$\alpha$+t($^{3}$He) cluster structure.
These two nuclei may be an example of the three-centre structure in which 
holes, rather than nucleons, are being exchanged between $\alpha$-cores.
The present measurements did not provide information on the spin and parity
of the observed states, which are unknown for some of the states, and also
on partial widths, which are crucial for the understanding of their structure.
Additional measurements capable of determining this information will be 
performed in the near future.

%\section{Summary}

\end{document}